\documentclass[useAMS,usenatbib,usegraphicx]{mn2e}

\title[New members of young moving groups]{A search for new members of the $\beta$\,Pic, Tuc-Hor and $\epsilon$\,Cha moving groups in the RAVE database}
\author[L.~L. Kiss et al.]{L.~L. Kiss$^{1,2}$\thanks{E-mail:
kiss@konkoly.hu}, A. Mo\'or$^{1}$,  T. Szalai$^{3}$, J. Kov\'acs$^{4}$, D. Bayliss$^{5}$, 
G. F. Gilmore$^{6}$, \newauthor O. Bienaym\'e$^{7}$, J. Binney$^{8}$,
J. Bland-Hawthorn$^{2}$, R. Campbell$^{9}$, K. C. Freeman$^{5}$, \newauthor J.P. Fulbright$^{10}$, B. K. Gibson$^{11}$, E. K. Grebel$^{12}$, A. Helmi$^{13}$,
U. Munari$^{14}$, \newauthor J. F. Navarro$^{15}$, Q. A. Parker$^{16,17}$, W. Reid$^{16}$, G. M. Seabroke$^{18}$, A. Siebert$^{7}$, \newauthor A. Siviero$^{14,19}$, 
M. Steinmetz$^{19}$, F. G. Watson$^{17}$, M. Williams$^{19}$, R. F. G. Wyse$^{10}$, \newauthor T. Zwitter$^{20,21}$\\
$^{1}$Konkoly Observatory of the Hungarian Academy of Sciences, PO Box 67, H-1525 Budapest, Hungary\\
$^{2}$Sydney Institute for Astronomy, School of Physics, University of Sydney, NSW 2006, Australia\\
$^{3}$Department of Optics and Quantum Electronics, University of Szeged, 6720 Szeged, D\'om t\'er 9., Hungary\\
$^{4}$Gothard Astrophysical Observatory, ELTE University, 9707 Szombathely, Hungary\\
$^{5}$Research School of Astronomy and Astrophysics, The Australian National University, Canberra, Australia\\
$^{6}$Institute of Astronomy, University of Cambridge, Madingley Road, Cambridge CB3 0HA, UK\\
$^{7}$Observatoire de Strasbourg, 11 Rue de L`Universit\'e, 67000 Strasbourg, France\\
$^{8}$Rudolf Pierls Center for Theoretical Physics, University of Oxford, 1 Keble Road, Oxford OX1 3NP, UK\\
$^{9}$Western Kentucky University, Bowling Green, Kentucky, USA\\
$^{10}$Johns Hopkins University, 3400 N Charles Street, Baltimore, MD 21218, USA\\
$^{11}$Jeremiah Horrocks Institute for Astrophysics \& Super-computing, University of Central Lancashire, Preston, UK\\
$^{12}$Astronomisches Rechen-Institut, Zentrum f\"ur Astronomie der
Universit\"at Heidelberg, D-69120 Heidelberg, Germany\\
$^{13}$Kapteyn Astronomical Institute, University of Groningen, Postbus 800, 9700 AV Groningen, Netherlands\\
$^{14}$INAF Osservatorio Astronomico di Padova, Via dell`Osservatorio 8, Asiago I-36012, Italy\\
$^{15}$University of Victoria, P.O. Box 3055, Station CSC, Victoria, BC V8W 3P6, Canada\\
$^{16}$Macquarie University, Sydney, NSW 2109, Australia\\
$^{17}$Australian Astronomical Observatory, P.O. Box 296, Epping, NSW 1710, Australia\\
$^{18}$Mullard Space Science Laboratory, University College
London, Holmbury St Mary, Dorking, RH5 6NT, UK\\
$^{19}$Astrophysikalisches Institut Potsdam, An der Sterwarte 16, D-14482 Potsdam, Germany\\
$^{20}$Faculty of Mathematics and Physics, University of Ljubljana, Jadranska 19, Ljubljana, Slovenia\\
$^{21}$Center of excellence SPACE-SI, Ljubljana, Slovenia\\
}
\begin{document}

\date{Accepted ... Received ...; in original form ...}

\pagerange{\pageref{firstpage}--\pageref{lastpage}} \pubyear{2010}

\maketitle

\label{firstpage}

\begin{abstract}
We report on the discovery of new members of nearby young moving groups, exploiting the full power of combining the RAVE survey with several stellar age diagnostic methods and follow-up high-resolution optical spectroscopy. The results include the identification of one new and five likely members of the $\beta$\,Pictoris moving group, ranging from spectral types  F9 to M4 with the majority being M dwarfs, one K7  likely member of the  $\epsilon$\,Cha group and two stars in the Tuc-Hor association. Based on the positive identifications we foreshadow a great potential of the RAVE database in progressing toward a full census of young moving groups in the solar neighbourhood. 
\end{abstract}


\section{Introduction} \label{intro}In the last decade many young ($<$100\,Myr) stars have been identified in the solar neighbourhood. 
Most of them belong to different moving groups, in which stars share common 
age and motion through the Galaxy \citep[for review see][]{zucksong2004}.
Up to now nine such kinematic assemblages have been revealed within 150\,pc of our Solar 
System with ages ranging from 8\,Myr to 70\,Myr \citep{torres2008}, most situated in the southern hemisphere
far from known star forming regions. However, tracing back their space trajectories shows that 
the birthplace of some of them may have been close to the nearest sites of massive star formation, 
the Sco-Cen region \citep{fernandez2008}.

Young moving groups are rich in debris discs, which implies
very active planetesimal formation around these stars \citep{moor2006,rebull2008}. 
Since the epochs of several key events in the early Solar System
\citep[e.g. formation of terrestrial planets,][]{apai2009} 
overlap well with the age of these groups, discs around the members 
are favourable and so present nearby and well-dated 
sites for investigations of planet formation and evolution. The members 
are also ideal targets when one would like to detect sub-stellar objects via direct imaging,
since giant gas planets are thought to fade significantly during their evolution 
\citep[e.g.][]{kasper2007}.

Most of the known young stellar kinematic groups occupy 
a large area on the sky (up to thousands of square degrees), which makes the identification of members very difficult.  
However, by combining astrometric data 
with radial velocity information and by applying relevant age diagnostic methods, one can search 
for additional members of known groups or reveal new 
kinematic assemblages. Using this approach, recent studies revealed more than 300 young stars
 belonging to nine kinematic groups in the vicinity of our Sun \citep{torres2008}. 
It is clear, however, that the census of these groups is far from complete, because of the lack 
of necessary kinematic information -- particularly the radial velocities and the trigonometric distances -- 
for most of the stars.
For example, comparing the list of members of the $\beta$\,Pic moving group with expectations based on the 
typical stellar mass function, \citet{shkolnik2009} estimated that about 60 M-type members remain undiscovered
in our neighbourhood.

The Radial Velocity Experiment (RAVE) is a large scale spectroscopic survey with the aim of measuring 
radial velocities and atmospheric parameters for up to a several hundred thousand stars in the southern sky, using the UK Schmidt telescope of the Anglo-Australian Observatory equipped with the 6dF multi-object spectrograph. The project has already resulted in two data releases \citep{steinmetz2006, zwitter2008}\footnote{See also {\tt http://www.rave-survey.org}} and observed well over 300,000 stars away from the plane of the Milky Way ($| b | > 25^{\circ}$) and with apparent magnitudes $9< I_{\rm DENIS}<13$. Detailed comparisons with external sets revealed that the velocities are accurate to 1-2 km~s$^{-1}$, while the errors of the stellar parameters are in the order of 400 K in temperature, 0.5 dex in gravity, and 0.2 dex in metallicity, all errors changing significantly across the temperature range of the stars \citep{zwitter2008}. Proper motion data in the catalogue have been taken from Tycho-2 for the brighter stars, then UCAC2, USNO-B and PPMX for the non-Tycho stars.

In order to capitalise on the potential of the RAVE survey in the census of young moving groups 
we used the 2008 August 30 internal release of the RAVE catalogue (with almost 250,000 entries) to search for new members 
of three young assemblages, the $\epsilon$\,Cha association, the $\beta$\,Pic moving group and the Tucana-Horologium association.
Each of the selected kinematic groups is younger than 40\,Myr.
From the location in the H-R diagram and from the lithium equivalent widths of the members 
\citet{zuckerman2001} derived an age of 12$^{+8}_{-4}$\,Myr for the $\beta$\,Pic moving group (BPMG).
Estimates based on dynamical back-tracing models of BPMG members, in good accordance with the previous value, 
yielded an age of $\sim$12\,Myr \citep{ortega2002,song2003}. Recently, \citet{mentuch2008} derived a somewhat 
higher age, 21$\pm$9\,Myr, from lithium depletion in the group.
For the slightly older Tucana-Horologium association (THA) the age estimates ranges between 10\,Myr and 40\,Myr 
\citep{zuckerman2001b,makarov2007,mentuch2008,daSilva2009}, at present the mostly approved value is 
30\,Myr \citep{torres2008}. 
The $\epsilon$\,Cha association may be the youngest among the three selected assemblages. 
Its age estimates range between 3\,Myr and 15\,Myr \citep{terranegra1999,feigelson2003,jilinski2005,torres2008}, 
most cases below 7\,Myr.
We used strict criteria to select potential candidates and then the RAVE-based list was supplemented by some 
additional stars taken primarily from the Hipparcos catalogue (Sect.~\ref{sampleselection}). 
In order to confirm the membership of our candidates we performed follow-up high-resolution spectroscopy (Sect.~\ref{obsanddatared}).
The final assignments of the candidate stars are summarised in Sect.~\ref{results}, with concluding remarks in Sect.~\ref{conclusion}.

\section{Sample selection} \label{sampleselection}



\subsection{Search in the RAVE catalogue}
The RAVE sample was examined in 
the U, V, W, X, Y, Z space, defined by the heliocentric space motion (U,V,W) and the physical space coordinates centred on 
the Sun (X,Y,Z). The computation of these parameters for a star requires knowledge 
of its coordinates, proper motion, radial velocity and distance. 
While coordinates (right ascension, declination), proper motion (in right ascension and 
in declination) and radial velocity information could be taken from the RAVE catalogue, 
for most of the RAVE stars no trigonometric distances were available. Therefore, 
the U,V,W,X,Y,Z values were calculated for a range of distances, between 5\,pc and 120\,pc with a resolution 
of 1\,pc, to check whether any distance resulted 
in a coordinate which coincided with the region of a specific association in this 6-dimension space.
In those cases where a RAVE star has a trigonometric parallax, measured by Hipparcos \citep{vanleeuwen07}, we used 
that value in the computations. 
The search was limited for those stars whose proper motion measurement fullfils the following criteria
1) $\rm \mu = \sqrt{ {\mu_{\alpha}}^2 {\cos{\delta}}^2 + {\mu_{\delta}}^2 } > 20$\,mas yr$^{-1}$; 2) 
$\rm {\mu}/\sigma_{\mu}>$5.

The characteristic space motion (U$_0$, V$_0$, W$_0$) of the groups were taken from \citet{torres2008}. 
We selected those objects from the RAVE catalogue, where $\min[((U-U_0)^2 + (V-V_0)^2 + (W-W_0)^2)^{1/2}] < 4$\,km~s$^{-1}$. The chosen limit corresponds to the internal dispersion of the known members of the groups, while the distance resulting the minimum value was adopted as the kinematic distance to the object. 
For the $\beta$ Pic moving group (hereafter BPMG), \citet{torres2006} found a correlation between the U component of the Galactic space motion 
and the X space coordinate that was taken into account by calculating the U$_0$ value as a function of X, using their Eq.\,4. 
We searched for BPMG candidate members within 80\,pc of the Sun.
For $\epsilon$\,Cha (ECA) and Tuc-Hor (THA) candidates the search was limited to a region defined by the known members of the groups
\citep[][]{torres2008}. Using this method we compiled an initial list of stars, that includes
3 ECA, 803 BPMG and 62 THA candidate members.

The initial lists were further evaluated and filtered: 
1) placing our candidates in the colour-magnitude diagrams of the specific kinematic groups; 
and 2) by searching for X-ray counterparts in the {\sl ROSAT} catalogues \citep{voges1999,voges2000}. 
We selected only those targets whose fractional X-ray luminosities (L$_{\rm x}$/L$_{\rm bol}$) 
and position in the colour-magnitude diagram were consistent with the similar properties of the known 
members (see Fig.~\ref{fig1}a--d). 
Our procedure finally resulted in 2, 9 and 7  
candidate members of the $\epsilon$\,Cha, BPMG and THA groups, respectively.
 By searching the literature we revealed that ten of our candidates  
(1 $\epsilon$\,Cha, 4 BPMG and 5 THA stars) are already known members. 
Moreover, one of our THA candidate, J042110.3-243221 (HD\,27679), has already 
assigned to the Columba moving group by \citet{torres2008}, while one of the BPMG candidate (TYC 7558-655-1) 
has an ambiguous assignment in the literature. \citet{torres2008} identified TYC 7558-655-1 as a possible 
member of the Columba group, on the other hand \citet{schlieder2010} proposed that this star likely belongs to the BPMG. 
We omitted these known/ambiguous members from the further observations and analyis. 
Thus, we finally selected six RAVE candidates (1 $\epsilon$\,Cha, 4 BPMG and 1 THA stars) for further investigations. 
These RAVE-based candidate list was supplemented by one additional star, J19560294-3207186, that is the comoving pair of one of 
new candidate object (TYC 7443-1102-1).

\subsection{Search in the HIPPARCOS catalogue}

 The Hipparcos catalogue was also searched for additional candidates members. 
Similarly to the RAVE sample, the Hipparcos stars were also examined in the U, V, W, X, Y, Z 
space. Here the radial velocity data are lacking for a significant fraction of stars, therefore the 
U, V, W values were calculated for a range of radial velocities 
(radial velocity values were varied between $-$50\,km~s$^{-1}$ and +50\,km~s$^{-1}$ with a resolution of 0.5\,km~s$^{-1}$).
For those stars where radial velocity data were available in the literature 
\citep{famaey2005,moor2006,pulkovo2006,torres2006,holmberg2007,kharchenko2007} we used the measured value 
in the computation. 
The search was limited to stars with spectral type later than F8. In the selection of candidate stars we applied 
almost identical criteria as in the case of RAVE objects. 
The only change in the method was related to those stars where all six parameters are available, thus U, V, W, X, Y, Z could be computed 
without any assumption, where we utilized a weaker criterion concerning to the X-ray luminosity of the object: 
we retained those candidates too where the upper limit of the X-ray luminosity was consistent with the similar property of the known members.
Using this method in the BPMG we could recover -- with the exception of HIP\,10679 -- all known members that are quoted in Hipparcos 
(and has spectral type F8 or later).
We note that HIP\,10679 composes a binary system with HIP\,10680 and the latter object has been successfully recovered by our method. 
Moreover, HIP\,10679 could be also recovered when we applied the more accurate trigonometric parallax of HIP\,10680 for this star as well. 
Three new candidate stars, HD\,37144, HD\,160305 and HD\,190102, have been revealed. For HD\,37144 and HD\,190102 radial velocity data and
lithium equivalent widths measured in the framework of the SACY survey \citep{torres2006} have already been available. 
Although the kinematic parameters fullfiled our criteria, the low values of the lithium equivalent widths did not confirm 
their membership. Based on similar considerations, \citet{daSilva2009} also rejected HD\,190102 as a member of BPMG. 
For HD\,160305 no radial velocity or lithium data were found in the literature, thus this object was added to the list of 
candidates. 
For THA and ECA we recovered all known members included in the Hipparcos catalogue. 
As a result of our search we revealed one new THA candidate, HD\,25402, for which radial velocity was available in the catalogue 
of \citet{holmberg2007}. Since its computed kinematic parameters correspond well to the characteristic values of THA we 
added this candidate to our list.


\begin{table*}
\setlength{\tabcolsep}{1.6mm}
\begin{center}
\scriptsize
\caption{Properties of the candidate stars. References for spectral types: 1 - \citet{riaz2006}; 2 - Hipparcos catalogue; 3 - 
\citet{lepine2009}; 4 - this paper, based on V$-$K$_{\rm s}$. For HD\,25402 and HD\,160305, distances were taken from the Hipparcos catalogue, otherwise we used kinematic distances (in parentheses). The typical uncertainty of the kinematic distances is estimated to be 
$\sim$10\%, based on a comparison between the kinematic and trigonometric distances of known members of BPMG and THA. The estimated uncertainty of the U,V,W components is about 1-2 km~s$^{-1}$.
\label{stellarprop}}
\begin{tabular}{lccccccccccc}
\hline\hline
Source ID                 &  RA (2000)    &  DEC (2000)  &  SpT. & V   &  $\rm K_s$  &   D   & v$_{\rm rad}$       &  U,V,W             &  EW$_{\rm Li}$ & EW$_{\rm H_{\alpha}}$ & $\log{\frac{L_{\rm x}}{L_{\rm bol}}}$ \\
                          &               &              &       &[mag]&   [mag]     &  [pc] &  [$\rm km\,s^{-1}$] & [$\rm km\,s^{-1}$] &  [$\rm \AA$]   &  [$\rm \AA$]          &           [dex]         \\
\hline
\multicolumn{12}{c}{candidate $\epsilon$ Cha group members} \\
J12210499-7116493  &  12 21 05.00  & -71 16 49.3 &   K7$^{1}$    &   12.16 &8.24     &  (98) &  +8.1$\pm$0.6      & [-11.8,-18.7,-9.8] & 0.550$\pm$0.020&-0.80$\pm$0.02  &-2.99   \\
\multicolumn{12}{c}{candidate $\beta$ Pic moving group members} \\
J01071194-1935359  &  01 07 11.94  & -19 35 36.0 &   M1$^{1}$    &   11.41 &7.25     &  (54) & +11.5$\pm$1.4      & [-8.6,-16.9,-8.3]  & 0.302$\pm$0.005&-2.00$\pm$0.05 &-3.13   \\
J16430128-1754274  &  16 43 01.33  & -17 54 26.9 &   M0.5$^{1}$  &   12.63 &8.55     &  (57) & -13.0$\pm$4.0      & [-11.3,-16.0,-6.6] & 0.300$\pm$0.020&-2.50$\pm$0.10 & -3.13   \\ 
HD160305           &  17 41 49.04  & -50 43 28.1 &   F9V$^{2}$   &   8.35 &6.99      &  72.5 &  +2.4$\pm$1.1      & [-6.1,-19.2,-10.5] & 0.130$\pm$0.040&2.60$\pm$0.40  & -3.67   \\ 
J19560294-3207186  &  19 56 02.94  & -32 07 18.7 &   M4$^{1}$    &   13.30 &8.11     &  (56) & -11.0$\pm$5.0      & [-9.8,-16.3,-8.1]  & $<$0.100       &-4.50$\pm$1.00 & -2.91 \\    
TYC 7443-1102-1    &  19 56 04.37  & -32 07 37.7 &   M0$^{3}$    &   11.80 &7.85     &  (56) &  -7.2$\pm$0.4      & [-9.8,-16.3,-8.1]  & 0.110$\pm$0.020&-0.68$\pm$0.04 &$<$-3.47 	\\
J20013718-3313139   &  20 01 37.18  & -33 13 14.0 &   M1$^{1}$    &   12.25 &8.24     &  (62) &  -5.6$\pm$1.8      & [-8.7,-16.5,-8.3]  & $<$0.100       &-1.03$\pm$0.07 &-3.39  \\
\multicolumn{12}{c}{candidate Tucana-Horologium association members} \\
J01521830-5950168  &  01 52 18.29  & -59 50 16.8 &   M2-3$^{4}$     &   12.94 &8.14     &  (39) &  +7.9$\pm$1.6      & [-10.2,-19.5,0.7] & $<$0.020        &-2.30$\pm$0.10 &-3.16   \\
HD25402            &  04 00 31.99  & -41 44 54.4 &   G3V$^{2}$   &    8.40 &6.88     & 48.5  & +16.3$\pm$0.7      & [-9.1,-20.9,-1.7] & 0.145$\pm$0.005 &+2.20$\pm$0.30 &$<$-4.02 \\
\hline
\end{tabular}  
\end{center}
\end{table*}

\section{Observations and data reduction} \label{obsanddatared}
We have obtained new high-resolution optical spectroscopy for all stars in Table~\ref{stellarprop} on six nights in 
July 2009 and three nights in August 2009, using the
2.3-m telescope and the Echelle spectrograph of the Australian National University.
The total integration time per object ranged from 30 s to 1800 s,
depending on the target brightness. The spectra covered the whole visual range in 27 echelle orders between 3900 \AA\ and 6720 \AA\, with only small
gaps between the three reddest orders. The nominal spectral resolution is $\lambda/\Delta \lambda\approx$ 23~000 at  the H$\alpha$ line, with typical
signal-to-noise ratios of about 100 (for the faintest red dwarf stars the blue parts of the spectra were much noisier).

All data were reduced with standard IRAF\footnote{IRAF is distributed by the National Optical Astronomy  Observatories, which are operated by the
Association of Universities for Research in Astronomy, Inc., under cooperative  agreement with the National Science Foundation.} tasks, including
bias and flat-field corrections, cosmic ray removal,  extraction of the 27 individual orders of the echelle spectra, wavelength calibration, and
continuum normalization. ThAr  spectral lamp exposures were regularly taken before and after every object spectrum to monitor the wavelength shifts
of  the spectra on the CCD. We also obtained spectra for the telluric standard HD~177724 and IAU radial velocity (RV) standards $\beta$~Vir (sp. type F9V)
and HD~223311 (K4III).

The spectroscopic data analysis consisted of two main steps. First, we measured radial velocities (RVs) by cross-correlating the target spectra (using the
IRAF task {\it fxcor}) with that of the RV standard that matched the spectral type of the target -- $\beta$~Vir was used for the early-type targets
(A--F--mid-G), HD~223311 for the late-type ones (late-G--K--M). Each spectral order was treated separately and the resulting velocities and the
estimated uncertainties were calculated as the means and the standard deviations of the velocities from the individual orders. For most of the
targets, the two IAU standards yielded RVs within 0.1--0.5 km~s$^{-1}$, which is an independent measure of the absolute uncertainties. 
Using the new, more accurate RV data we recomputed the U, V, W, X, Y, Z values for each candidate star, as described earlier. The equivalent width of the 6708\AA\, Li and the H$_{\alpha}$ lines were measured with the 
IRAF task {\sl splot}.

\section{Results} \label{results}

Though stars in a specific group can be widely scattered across the sky, 
their common properties offer an opportunity to identify other members in the field by prescribing
that a candidate must have similar space motion, as well as age, to the known members. 
The age criterion is essential because there may be a non-negligible fraction of old 
field stars which have similar space motion to the young group members \citep{zucksong2004,lopez2009}.

Using the new RV data we recomputed the U, V, W values for each candidate star and compared the values 
with the characteristic space motion (U$_0$, V$_0$, W$_0$) of the corresponding kinematic groups. 
Apart from HD\,25402 and HD\,160305, we used kinematic distances and the proper motion data were taken from the 
UCAC3 catalogue \citep{zacharias2010} for all of our targets. 
For HD\,25402 and HD\,160305, distances and proper motions were taken from the Hipparcos catalogue.
For J19560294-3207186,
 we adopted the kinematic distance and U, V, W values of its companion (TYC 7443-1102-1). 
We required $[((U-U_0)^2 + (V-V_0)^2 + (W-W_0)^2)^{1/2}] < 4$\,km~s$^{-1}$ for the group membership. 
This criterion was fulfilled for all candidate stars, i.e. none of the refined RV data resulted in a deprived 
candidacy. 

We used three different age diagnostic methods to evaluate whether the candidates are approximately coeval 
with the corresponding kinematic groups. In all age indicators we compared the specific properties of  
the candidates stars to the corresponding properties of the known members with similar colour indices.

\begin{figure*} 
\includegraphics[width=170mm]{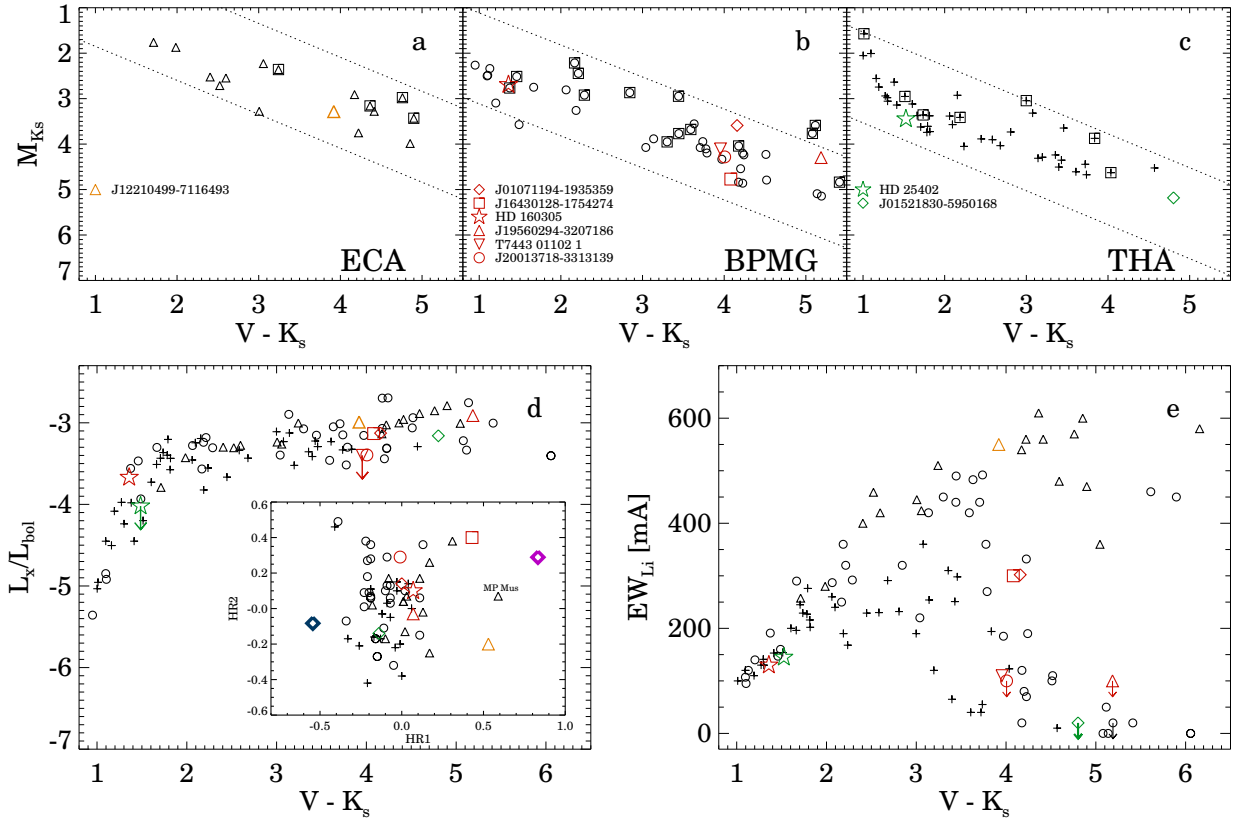}
\caption{ {\sl Upper panels:} Absolute K$_{\rm s}$ magnitude versus V$-$K$_{\rm s}$ colour diagrams for the known and candidates members of ECA 
({\sl panel a}), BPMG ({\sl panel b}) and THA ({\sl panel c}) kinematic groups. 
Following \citet{lepine2009}, the loci of the known members of the specific groups
were fitted by a line, the $\pm$1 magnitude range of the fitted loci are denoted by dotted lines in a--c panels.
Nearly all known members are located within these lines in the diagrams.
Candidate objects out of these area were omitted from our survey (see Sect.~\ref{sampleselection}).  
{\sl panel d:} Fractional X-ray luminosities as a function of V$-$K$_{\rm s}$ for the known and candidates members of the three kinematic groups.
The inset shows the X-ray hardness ratio HR1 vs. hardness ratio HR2 for the same objects.  Large purple and blue diamonds show 
the characteristic hardness values for T\,Tauri stars and old K- and M-type main-sequence 
field stars, respectively. The characteristic values were taken from \citet{kastner2003}.                                   
The inset does not cover any symbol of the large panel. 
{\sl panel e:} Equivalent width of Li~$\lambda$6707.76 as a function of V$-$K$_{\rm s}$. 
In all panels known members of BPMG, THA and ECA are plotted as black circles, black plus signs and black triangles, respectively. 
Squares denote those known close binaries where no photometric data are available separately for the individual components.   
Upper limits are displayed with down arrows.
}
\label{fig1}
\end{figure*}

Figure~\ref{fig1}\,a,b,c show the colour-magnitude diagrams (M$_{\rm Ks}$ vs. V$-$K$_{\rm s}$) of the ECA, BPMG and 
THA groups. The lists of known group members are from \citet{torres2008}, 
which have then been queried in the Hipparcos and 2MASS databases to produce
Fig.~\ref{fig1}\,a,b,c. Candidate members are plotted with different symbols in the 
corresponding panels, showing that they indeed occupy a distinct region in the CMD. 
Since in these young groups a significant fraction of the stars are in pre-main 
sequence evolutionary stage, the characteristic loci of the members in the CMD deviate from the position 
of the main-sequence stars, which helps filter out 
spurious (old) candidates. The younger the association the higher the deviation, because more and more 
massive stars are still in pre-main sequence stage.

Young stars are also known to have enhanced coronal activity with 
strong X-ray emission making the latter property a good indicator of youth.  
We have cross-correlated the list of the candidates and known members of the three groups with 
the ROSAT All-Sky Survey catalogues \citep{voges1999,voges2000}. We selected only those objects where the 
match between the optical and X-ray positions was within 40{\arcsec}. 
In all of the positive matches we checked the DSS images 
to evaluate whether there are any other nearby sources of X-ray emission within the ROSAT positional
uncertainties. The X-ray fluxes of the sources were computed using the count rate to-energy flux conversion formula 
by Schmitt et al. (1995).
For those two objects where no X-ray counterparts were found we utilized the ROSAT All-Sky survey 
images to derive an upper limit in the X-ray flux.
Figure~\ref{fig1}d displays the fractional X-ray luminosities vs. V$-$K$_{\rm s}$ for 
the group members and the candidate stars.  
The ROSAT X-ray hardness ratios (HR1 and HR2) of the X-ray counterparts 
were also plotted in an inset of Fig.\ref{fig1}d.
Analyzing the ROSAT hardness ratio values (HR1 and HR2) for T\,Tauri stars, young moving group members and 
for old field stars, \citet{kastner2003} demonstrated that the X-ray spectra of F through M stars 
soften with age. They argued that the distributions of hardness ratios for BPMG, THA and TW Hya association members 
are tightly clustered and very similar to one another.
Thus these ratios can also be used to discriminate between young and old stars.

Lithium is burned at low temperatures (2.5$\times$10$^6$\,K) in stellar interiors. Since lithium is destroyed and never created
in nuclear reactions, primordial lithium depletes monotonically with time in stars with a convective layer. It makes lithium one of the best age indicators for young stars \citep{zucksong2004}. 
The measured lithium equivalent widths of the candidates (from Table~\ref{stellarprop}), as well as the known group members are plotted in 
Figure~\ref{fig1}e as a function of V$-$K$_{\rm s}$. 
For known group members the lithium data were taken from different 
surveys \citep{torres2006,daSilva2009}.

In tight binaries the tidal interactions can induce strong activity indicating a spurious 
young age even if the system is old otherwise. Thus in these cases the usage of 
X-ray luminosity as a diagnostic of youth could be erroneous. 
The confusion between single and binary stars can also lead to the estimates of spurious stellar parameters.
Our high-resolution spectra allowed us to search for double- and multilined binaries by 
cross-correlating with the IAU RV standards. We found no double-lined binaries among our sample stars.  
Apart from HD\,160305 -- where no previous RV data were available -- velocities measured by us all agree 
well with those in the RAVE data or published by \citet{holmberg2007}.
Thus currently there is no evidence to suggest that any of our candidates reside in a tight multiple system.

\subsection{Final assignment of candidate stars}

Unfortunately, we have parallax information only for two of our candidates (HD\,25402 and HD\,160305). 
The lack of reliable trigonometric distances for RAVE stars makes the identification of new members ambiguous,  
even in those cases where the age diagnostic methods confirm the youth of the objects, because we 
cannot completely exclude the possibility that we observe a young star whose kinematic properties deviate from those of the 
moving group because its real distance deviate from the one we predicted using our method. 
Thus the RAVE-based candidates  -- in accordance with the nomenclature proposed by \citet{schlieder2010} -- 
are classified as likely new members even in those cases when all of the prescribed membership criteria are fulfilled.

{\sl J12210499-7116493:} Its position coincides well with the known members of 
ECA in all age diagnostic figures (Fig.~\ref{fig1}\,a,d,e). The star has a somewhat larger HR1 ratio than most of  
the young group members, similar to the classical T\,Tauri star MP\,Mus recently assigned to ECA 
by \citet{torres2008}. \citet{kastner2003} proposed that for T\,Tauri stars the presence of star-disc interaction and especially 
accretion can explain the stronger X-ray hardness ratios.
Using the empirical criterion proposed by \citet{barrado2003} to distinguish between stars with chromospheric activity and objects with 
accretion we conclude that the weak H$_\alpha$ emission of J12210499-7116493 (see Table~\ref{stellarprop}) may be of 
chromospheric origin. 
Based on photometric data from ASAS3, \citet{bernhard2009} found J12210499-7116493 to be a probable BY\,Dra type variable 
star with a period of 6.855\,days. 
Since the properties of J12210499-7116493 fulfilled all of our criteria we propose it is a likely new member of ECA.

{\sl J01071194-1935359 and J16430128-1754274:}
All three age determination methods confirm that these stars are likely to be coeval with the BPMG.  
Although their spatial location somewhat deviate from the location of the known members, we classify both stars as likely new members of the BPMG. 

{\sl HD\,160305:} 
This star is located quite close to three other BPMG stars (GSC 8350-1924, CD-54 7336, HD\,161460), 
all within a sphere of 14\,pc across. The position of the star in the age diagnostic diagrams 
is in good accordance with the known members of BPMG, hence we identify HD\,160305 as a new member of the BPMG.

{\sl J19560294-3207186, TYC 7443-1102-1  and J20013718-3313139:} 
\citet{lepine2009} proposed that TYC 7443-1102-1 and J19560294-3207186, forming a common proper motion pair, belong to the BPMG. Our observations showed that the radial velocities of the two stars are also in good agreement within the uncertainty 
of the measurements. J20013718-3313139, one of our candidates, is located at an angular distance of
1.6{\degr} from  TYC 7443-1102-1/J19560294-3207186.
Both the proper motions ($\mu_{\alpha}\cos{\delta}=27.1\pm2.6$\,mas, $\mu_{\delta}=-60.9\pm1.8$\,mas) 
and the radial velocities ($v_{rad}=-5.6\pm1.8$\,km\,s$^{-1}$) of J20013718-3313139 are in good agreement 
with the corresponding astrometric properties of TYC 7443-1102-1 
($\mu_{\alpha}\cos{\delta}=30.3\pm1.5$\,mas, $\mu_{\delta}=-66.6\pm1.0$\,mas, $v_{rad}=-7.2\pm0.4$\,km\,s$^{-1}$). 
The derived kinematic distance of J19560294-3207186 is also close to that of TYC 7443-1102-1. 
We note that all three stars overlap well with the space distribution of previously known BPMG members. 
Even if J20013718-3313139 and TYC 7443-1102-1/J19560294-3207186 do not form a bound system,
the stars are likely to be co-eval, therefore we can combine the results of different age diagnostic methods for the three objects.
The absolute magnitudes and the X-ray fractional luminosities (only an upper limit for TYC 7443-1102-1) overlap well with the locus of known BPMG members in the CMD (Fig~\ref{fig1}\,b) and in X-ray related diagrams (Fig~\ref{fig1}\,d).
On the other hand, the Li equivalent width of TYC 7443-1102-1 and J20013718-3313139 (the latter is only an 
upper limit) are somewhat lower 
than that of the known members with similar V$-$K$_s$ colour.
We classified these three stars as likely new BPMG members.

{\sl J01521830-5950168:}
For stars with V$-$K$_{\rm s} > $4.5 in the THA, lithium is burning on a timescale shorter than the age of the group ($\sim$30\,Myr). 
Thus, this age diagnostic method cannot be used to support or reject the youth of J01521830-5950168.
Since both the X-ray properties and the position of the star in the CMD coincide well with the known THA members, 
we list this object as a likely new member of THA.

{\sl HD\,25402:}
Both the K$_{\rm s}$ absolute magnitude and the measured lithium equivalent width of HD\,25402 are consistent with the 
similar properties of early G-type stars of the THA. The star was not detected in the ROSAT survey.
HD\,25402  is identified as a wide binary system in the CCDM catalogue \citep{dommanget2002}. 
However, according to the UCAC3 catalogue \citep{zacharias2010}, 
the proper motion of the proposed companion (CCDM J04005-4145B, 
$\mu_{\alpha}\cos{\delta}=60.2\pm1.7$\,mas, $\mu_{\delta}=-32.4\pm1.8$\,mas ) deviates from the proper motion of 
HD\,25402 ($\mu_{\alpha}\cos{\delta}=70.2\pm0.9$\,mas, $\mu_{\delta}=+2.0\pm2.1$\,mas), implying that a physical connection between these stars is unlikely.
Based on the current data we assign this star to the THA. 

\section{Conclusion} \label{conclusion} 

We searched for new members of three young kinematic assemblages ($\beta$\,Pic moving group, 
$\epsilon$\,Cha and Tucana-Horologium associations) by combining radial velocity data 
from the RAVE survey with other astrometric information. We used strict selection criteria to filter out 
false candidates by requiring consistency with the colour-magnitude relationship and X-ray properties 
of the known members. In addition to recovering 10 known members of the three groups, we identified 
seven late-type (K,M) new candidates. This list was supplemented by two additional member candidates, HD\,25402 and HD\,160305, 
that were selected from the Hipparcos catalogue.

Utilizing our new high-resolution spectroscopic observations we found further pieces of evidence for the membership of our targets. As a result, we identified two new members (1 BPMG, 1 THA) and seven 
likely  members ( 1 ECA, 5 BPMG, 1 THA) of the groups.  All stars, except TYC 7443-1102-1 and J19560294-3207186, two likely BPMG members identified by \cite{lepine2009}, are new discoveries.     
These results demonstrate the potential of the RAVE survey in improving the census of young moving groups. Using the same methods 
the searches can be extended for other moving groups as well. Moreover, the final version of the RAVE catalogue will contain 
data for approximately three times more stars, offering a great opportunity for further steps towards a full census 
of young kinematic groups in the Galactic neighbourhood.

\section*{Acknowledgments}
This project has been supported by the Australian Research Council, the University of Sydney, the 'Lend\"ulet' Young Researchers 
Program of the Hungarian Academy of Sciences, and the Hungarian OTKA Grants K76816, MB0C 81013 and K81966. Funding for RAVE  has been provided by the  Anglo-Australian Observatory, by
the Astrophysical Institute Potsdam,  by the Australian Research Council, by
the German  Research foundation, by the National  Institute for Astrophysics
at  Padova, by  The Johns  Hopkins University,  by the  Netherlands Research
School  for Astronomy,  by  the Natural  Sciences  and Engineering  Research
Council of Canada,  by the Slovenian Research Agency,  by the Swiss National
Science  Foundation,   by  the  National  Science  Foundation   of  the  USA
(AST-0508996), by  the Netherlands Organisation for  Scientific Research, by
the Particle Physics  and Astronomy Research Council of  the UK, by Opticon,
by Strasbourg Observatory, and by  the Universities of Basel, Cambridge, and
Groningen. The RAVE web site is at www.rave-survey.org.

\label{lastpage}

\end{document}